\documentclass[useAMS,usenatbib,usegraphicx]{mn2e}

\title[Fading Features Found in the Galactic Stellar Halo] {Fading Features Found in the Kinematics of the Far-Reaching Milky Way Stellar Halo}
  \author[Sarah A. Bird and Chris Flynn]{Sarah~A.~Bird$^{1,}$$^{2}$
  \thanks{E-mail: sarah.bird@utu.fi} 
  and 
  Chris~Flynn$^{3}$\\
  $^{1}$Key Laboratory for Research in Galaxies and Cosmology, 
  Shanghai Astronomical Observatory, Chinese Academy of Sciences,\\ 
  80 Nandan Road, Shanghai, 200030, China\\
  $^{2}$Tuorla Observatory, Department of Physics and Astronomy, 
  University of Turku, V\"ais\"al\"antie 20, FI-21500 Kaarina, Finland\\
  $^{3}$Centre for Astrophysics and Supercomputing, 
  Swinburne University of Technology, Hawthorn, VIC, 3122, Australia
}
\begin{document}

\date{Accepted 2015 June 26. Received 2015 June 25; in original form 2014 December 2}

\pagerange{\pageref{firstpage}--\pageref{lastpage}} \pubyear{2015}

\maketitle

\label{firstpage}

\begin{abstract}
We test the long-term kinematical stability of a Galactic stellar halo model,
due to \citet{Kafle2012}, who study the kinematics of approximately 5000 blue
horizontal branch (BHB) stars in the Sloan Digital Sky Survey (SDSS). The velocity
dispersion $\sigma$ and anisotropy parameter $\beta$ of the stars have been
determined as functions of Galactocentric radius, over the range $6< 
R_\mathrm{GC} < 25$ kpc, and show a strong dip in the anisotropy profile at
$R_\mathrm{GC}\sim17$ kpc. By directly integrating orbits of particles in a 3-D model
of the Galactic potential with these characteristics, we show that the
$\sigma$ and $\beta$ profiles quickly evolve on a time scale of a
$\mathrm{few}\times10$~Myr whereas the density $\rho$ profile remains largely
unaffected.  We suggest that the feature is therefore transient. The origin of
such features in the Galactic halo remains unclear.

\end{abstract}


\section{Introduction}

Studying the Milky Way's stellar halo is an important route to understanding
galaxy formation, as the halo is such an old Galactic component. 
Intrinsically
bright stars with easily measured radial velocities have been the usual means
of doing so, with red giants and horizontal branch stars as typical tracers in such 
studies. Early studies of the stellar halo kinematics date to the 1950s, and
focused on halo stars passing through the Solar neighbourhood, but it was not
until the 1980s that large $(\ga 100)$ samples of halo stars tens of kpc from
the Sun began to be collected and analysed (see the reviews by \citet{Sandage1986} and \citet{Helmi2008}).

Milky Way halo BHB stars from $\sim5$ to 50 kpc have
been studied by \citet{Sommer-Larsen1994}. They used about 100 stars to develop
a kinematical model of the outer Milky Way halo, with the surprising result
that the orbits of stars in the far outer halo ($> 20$ kpc) appear to be much more
tangential than radial. \citet{Flynn1996} used simulations of such stars
orbiting in the Milky Way potential which showed such a distribution of halo
orbits is stable over a Hubble time.

Since then, numerous studies have added to the sample of BHB halo stars
\citep{Sommer-Larsen1997,Sirko2004,Deason2011,Deason2012} but show a wide
spread in the resulting kinematical models for the outer stellar halo.  

\citet{Sommer-Larsen1997} analysed about 700 BHB stars, mainly within
20 kpc of the Sun, but also probing out to 50 kpc. They found that the
outer stellar halo velocity dispersion (at $\approx 50$ kpc) was
quite ``cold'' (i.e. low velocity dispersion), nearing 100 km s$^{-1}$ compared with the value at the sun $\simeq140$ km~s$^{-1}$. They concluded that outer halo orbits must be
quite tangential (with a tangential velocity dispersion of about 150
km s$^{-1}$), given the observed density distribution of halo stars and
assumptions about the Milky Way's dark matter distribution.

On the other hand, \citet{Sirko2004} have advocated an isothermal
outer halo ($R_\mathrm{GC} \ga R_\odot$), in which all three components of the
velocity dispersion are $\approx 100 $ km s$^{-1}$, based on $\approx 1200$
BHB stars from SDSS.  \citet{Thom2005}
subsequently analysed 530 BHB stars with radial velocities and
distances from the Hamburg/ESO survey, finding it difficult to
discriminate between the simplest, isothermal kinematic models and
anything more complex, and advocating further studies of the inner
halo to help resolve the issue.

Very distant BHB stars have recently been shown by \citet{Deason2011} and \citet{Deason2012}
to have very ``cold'' kinematics -- low velocity dispersions of
$\approx 50-60$ km s$^{-1}$ in the radial range 100 to 150 kpc. The density
falloff in these regions is much steeper than inside 100 kpc, and the
dynamical times rather long, so the region is unlikely to be well
mixed. Comparison with the kinematical models of the inner region,
which implicitly assume the halo is well mixed, are thus difficult.

\begin{figure}
\centering \includegraphics[width=8cm]{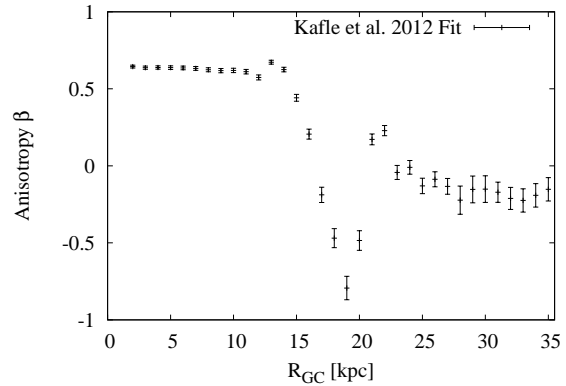}
\caption{Anisotropy $\beta$ profile as a function of Galactocentric radius,
  $R_\mathrm{GC}$, showing $\beta$ as calculated from the input
  \citet{Kafle2012} radial and tangential velocity dispersions, $\sigma_r$ and
  $\sigma_t$. \citet{Kafle2012} measured the anomalous feature at its extremum
  to be $\beta=-1.2$ at $R_\mathrm{GC}=17$ kpc. The error bars are Poissonian.
}
\label{beta}
\end{figure}

More recently, \citet{Kafle2012} have used
$\approx5000$ BHB halo stars found in the SDSS/Sloan Extension for Galactic
Understanding and Exploration (SEGUE) to analyse the kinematics of the Galactic
stellar halo. Working from photometric
distance estimates and radial velocity measurements for each star, they performed a 
maximum likelihood analysis to determine the
velocity dispersion $\sigma$ and anisotropy $\beta$ profile, where $\beta$ is
defined by \citet{Binney2008} in spherical coordinates using radial $\sigma_r$ and tangential 
($\sigma_\theta, \sigma_\phi$) velocity dispersions such that
\begin{equation}
\beta = 1 -\frac{\sigma_\theta^2 + \sigma_\phi^2}{2\sigma_r^2}.
\end{equation}
\citet{Kafle2012} found
a previously unseen
feature, most prominent in the $\beta$ profile measured out 
to a Galactocentric radius of $R_\mathrm{GC}\approx25$ kpc, which shows a rapid
decline at $R_\mathrm{GC} = 13$ kpc, reaching a minimum at $R_\mathrm{GC} =
17$ kpc, followed by a sharp rise within just a few kpc (Fig.~\ref{beta}). 
This feature has been confirmed
in a very recent study of the halo with an even larger sample of stars by \citet{KingarXiv2015}.

Given that the feature is so narrow and deep, but is made up of stars in an
otherwise ``hot'' (i.e. high velocity dispersion) Galactic halo, we were
motivated by the question of whether such a feature could be long-term stable
or simply transient.

We have tested this by setting up simulations of the stars in the Milky Way
potential, and determined the model's stability over the order of a Hubble
time, finding that it is transient, and dissolves away in just a few tens of
Myr.

In Sec.~\ref{model} we describe our choice of potential to use for our
simulations (\citet{Flynn1996}), updating it to be consistent with the
remarkable new constraints on the Galactic potential by \citet{Bovy2013}.
In Sec.~\ref{sims}, we describe our simulations testing stability over time of
the stellar halo density distribution $\rho$, the velocity dispersion $\sigma$,
and anisotropy $\beta$ profiles, motivated as such by the analysis of
\citet{Kafle2012}. We discuss our results and draw conclusions in
Sec.~\ref{conclusions}.

\section{Galactic Potential Model}
\label{model}

We used a Milky Way potential model similar to that of \citet{Flynn1996}.  The
potential consisted of the sum of
three components, namely the the potential due to the dark halo, a central
component, and the disc. The dark halo potential was spherical with
mass of order $10^{12}\mathrm{M}_\odot$ within 100 kpc. The central
potential was modeled by the sum of a spherical bulge/stellar-halo and a
spherical inner core potential. The disc potential itself consisted of three
Miyamoto-Nagai potentials \citep{MiyamotoNagai1975}. In this model, the disc
has a scalelength of $R_\mathrm{D}=2.2$ kpc, to be consistent with the
measurements of \citet{Bovy2013} from the kinematics of over 16,000 G-type
dwarfs in the SDSS/SEGUE survey distributed between Galactocentric radii of
$5<R_\mathrm{GC}/\mathrm{kpc}<12$. Our Galactic potential model parameters are
listed in Table \ref{parameters}. The parameters have been set such that the
rotation curve, i.e. circular velocity as a function of Galactocentric radius,
is flat out to $R_\mathrm{GC}=500~\mathrm{kpc}~-$ i.e. the outer limit of our
modeled halo (although we only analyse the stars within 100 kpc). We adopted
the same parameters used by \citet{Kafle2012}, namely the Galactocentric
position of the Sun at $R_\odot = 8.5$ kpc and the velocity of the local
standard of rest as $v_\mathrm{LSR} = 220$ km s$^{-1}$. 

\begin{table}
\caption{Adopted Parameters for the Galactic Potential}
\label{parameters}
\begin{center}
\begin{tabular}{cccc}
\hline
Component & Parameter & Value & Unit\\
\hline
Dark Halo & $r_\mathrm{H}$ & 15.0 & kpc\\
 & $M (R_\mathrm{GC}\la 100~\mathrm{kpc})$ & $\sim10^{12}$ & M$_\odot$\\
 & $V_\mathrm{H}$ & 220 & km s$^{-1}$\\
\hline
Bulge/Stellar Halo & $r_\mathrm{C_1}$ & 2.70 & kpc\\
 & $M_\mathrm{C_1}$ & 3.0 & $10^9\mathrm{M}_\odot$\\
Central Comp. & $r_\mathrm{C_2}$ & 0.42 & kpc\\
 & $M_\mathrm{C_1}$ & 16 & $10^9\mathrm{M}_\odot$\\
\hline
Disc & $b$ & 0.3 & kpc\\
 & $r_\mathrm{D_1}$ & 5.81 & kpc\\
 & $M_\mathrm{D_1}$ & 106 & $10^9\mathrm{M}_\odot$\\
 & $r_\mathrm{D_2}$ & 17.43 & kpc\\
 & $M_\mathrm{D_2}$ & $-45.8$ & $10^9\mathrm{M}_\odot$\\
 & $r_\mathrm{D_3}$ & 34.86 & kpc\\
 & $M_\mathrm{D_3}$ & 5.24 & $10^9\mathrm{M}_\odot$\\
\hline
\end{tabular}
\end{center}
\end{table}


\section{Numerical Methods and Simulations}
\label{sims}

\begin{figure}
\centering \includegraphics[width=7cm]{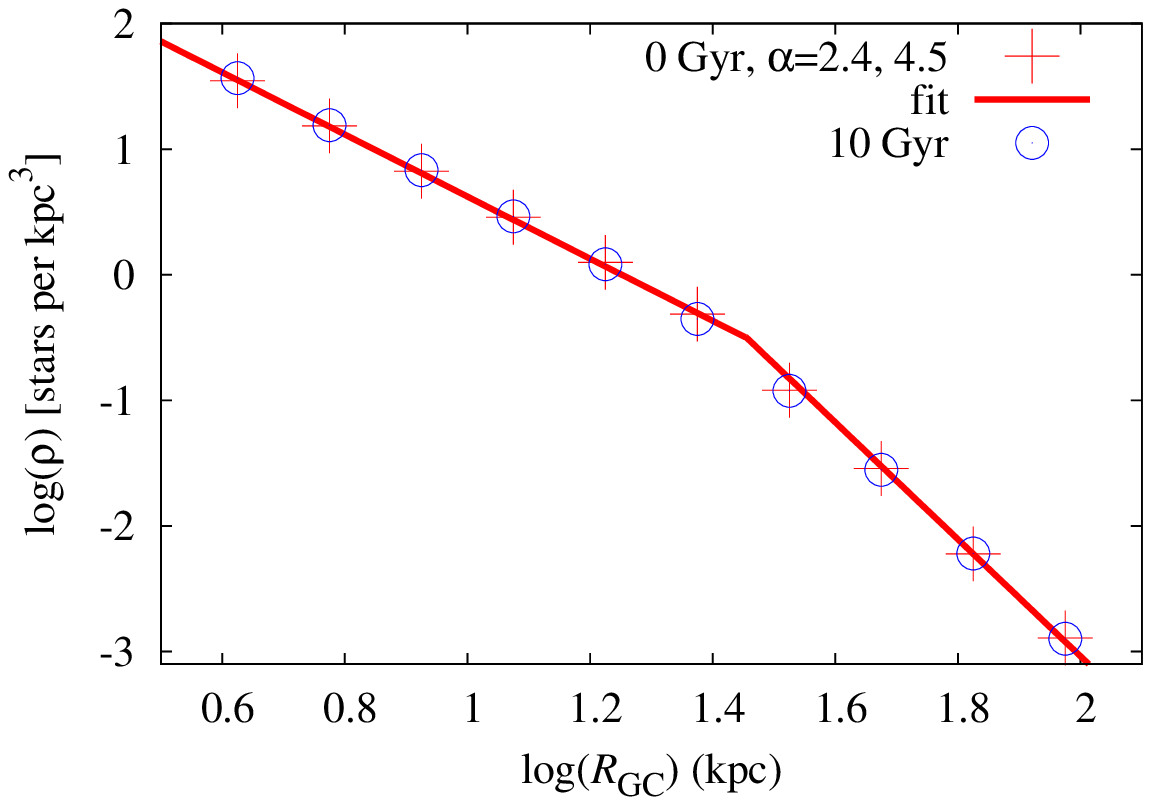}
\caption{Radial density $\rho$ profile of simulated halo stars. The initial
  density at 0 Gyr is the broken power law assumed by \citet{Kafle2012} (as
  observed by \citet{Watkins2009} and \citet{Deason2011}), where $\rho\propto
  R_\mathrm{GC}^{-\alpha}$ and $\alpha=2.4$ at $R_\mathrm{GC}\leq27$~kpc and
  $\alpha=4.5$ at $R_\mathrm{GC}>27$~kpc (red crosses with red fit line). The
  profile is plotted at 0 and 10 Gyr (blue circles), showing the stability of
  $\rho$ as we see negligible change between the initial and final states. The
  break radius at which the two power laws meet is $R_\mathrm{GC}\approx27$
  kpc, or $\log(R_\mathrm{GC})\approx1.43$ kpc.  }
\label{hist}
\end{figure}
\begin{figure}
\centering \includegraphics[width=7cm]{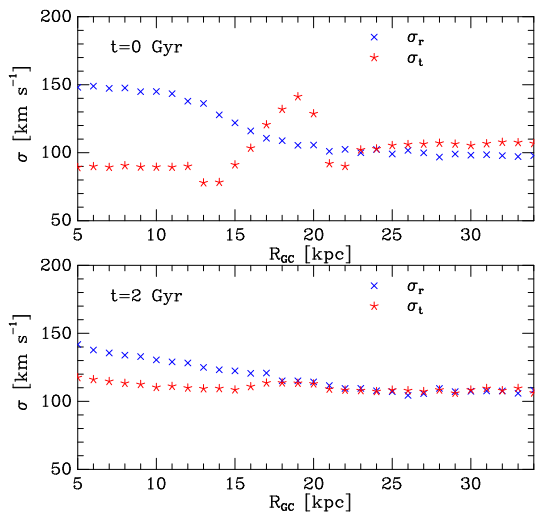}
\caption{Upper panel displays the initial velocity dispersion $\sigma$ profile
  from \citet{Kafle2012}. Radial and tangential velocity dispersions,
  $\sigma_r$ (blue crosses) and $\sigma_t$ (red asterisks) (km s$^{-1}$),
  respectively, are plotted as a function of distance from the galactic centre,
  $R_\mathrm{GC}$ (kpc).  The lower panel shows the $\sigma$ profiles after 2
  Gyr. The initial turnover features at $R_\mathrm{GC}\approx17$~kpc in the
  model halo disappear within 0.02 Gyr and the new $\sigma$ profiles remain
  stable in the simulations.  }
\label{veldisp_initial}
\end{figure}
We tested for stability over time of the Galactic stellar halo density
distribution $\rho$, velocity dispersion $\sigma$, and anisotropy $\beta$
profiles, setting them up to initially match the observations of
\cite{Kafle2012}. We ran simulations for 10~Gyr using $2.24\times10^5$
particles in the \citet{Flynn1996} Galactic potential as described in
Sec. \ref{model}. The stellar velocities were initially randomly drawn from
Gaussian distributions and set up in a spherical non-rotating
configuration. The density profile (Fig.~\ref{hist}) followed the double
power law assumed by \citet{Kafle2012}, as observed by \citet{Watkins2009} and
\citet{Deason2011} where $\rho\propto R_\mathrm{GC}^{-\alpha}$ with
$\alpha=2.4$ at $R_\mathrm{GC}\leq27$~kpc and $\alpha=4.5$ at
$R_\mathrm{GC}>27$~kpc (red crosses with red fit line). Very interestingly, \citet{Deason2014} found recently a striking drop in the stellar density profile, with $\alpha=6$ at 50 kpc and $\alpha=6-10$ out to 100 kpc. Although the stellar distribution of our simulations was truncated at 500 kpc and 0.1 kpc, we do not include the most recently found drop in stellar density as it would have negligible effects on the stability of the system at much smaller radii near 17 kpc.

The initial velocity dispersions profiles are shown in the upper panel of
Fig. \ref{veldisp_initial}.

Our simulations are of tracer stars in a fixed Galactic potential and the 
stars did not interact. The stellar orbits were integrated using a Runge-Kutta scheme control
\citep{Press1986} with a 0.01 Gyr (adaptive) stepsize over a period of 10 Gyr,
and the density $\rho$, velocity dispersion $\sigma$, and anisotropy $\beta$
profiles of the particles as functions of $R_\mathrm{GC}$ were determined at 2
Gyr intervals.

\begin{figure*}
\centering \includegraphics[width=15cm]{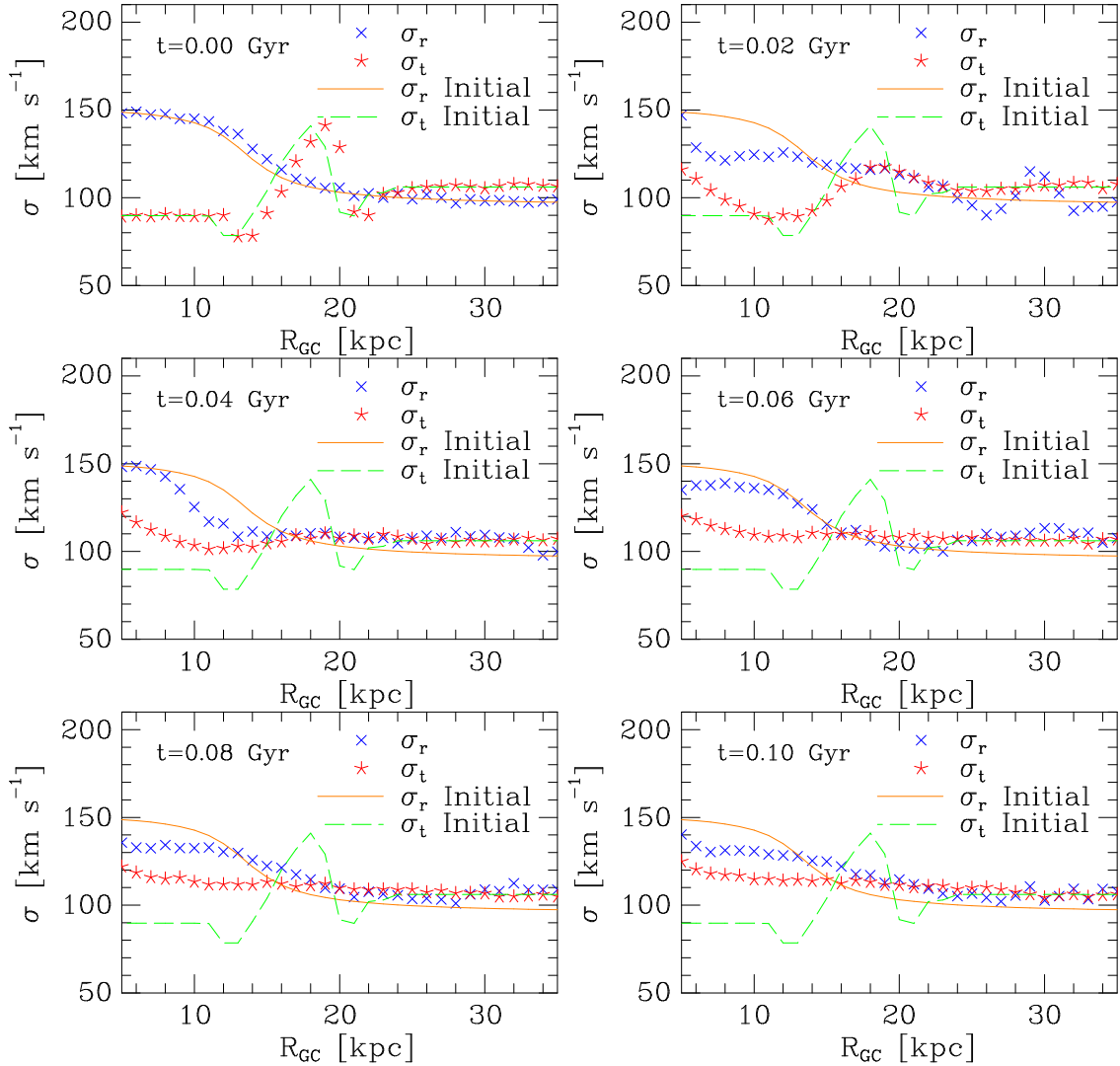}
\caption{Similar to Fig.~\ref{veldisp_initial}, but shown over the first 0.1
  Gyr in the simulation. Radial and tangential velocity dispersions, $\sigma_r$
  (blue crosses) and $\sigma_t$ (red asterisks) (km s$^{-1}$), are plotted as a
  function of distance from the Galactic centre, $R_\mathrm{GC}$ (kpc). The
  $\sigma$ profiles are plotted every 0.02 Gyr as indicated in the top left
  corner of each panel. The model input $\sigma$ profiles, as observed by
  \citet{Kafle2012}, are plotted in each panel as the orange line ($\sigma_r$)
  and the green dashed line ($\sigma_t$). The $\sigma_r$ and $\sigma_t$
  profiles change within the first 0.02 Gyr (top right panel). After 0.1 Gyr,
  the profiles have relaxed (lower right panel) and change negligibly over the
  remainder of the simulation.  }
\label{veldisp_Myr}
\end{figure*}

We found that the $\rho$ profile was stable over the 10 Gyr simulation, as seen
in Fig.~\ref{hist} when comparing the negligible changes between the initial
(red crosses) and final (blue circles) profiles.  Unlike the stable $\rho$
profile, the initial \citet{Kafle2012} $\sigma$ profiles as seen in the top
panel of Fig.~\ref{veldisp_initial} quickly relaxed to a new state, appearing
in the lower panel of Fig.~\ref{veldisp_initial}, which was stable over the
remaining simulation time. Fig.~\ref{veldisp_Myr} shows that the initial
profile was quickly changed within the first $0.02$ Gyr (top right panel of
Fig.~\ref{veldisp_Myr}). By 0.1 Gyr (lower right panel of \ref{veldisp_Myr}),
relaxation had occurred and the new $\sigma$ profiles were stable over the
remaining 8 Gyr.  Similarly, the initial feature in $\beta$ disappeared and the
profile remained stable during the rest of the simulation. In addition, we tested variations in the initial conditions which were consistent with the boundaries set by the error bars of the \citet{Kafle2012} determinations of sigma versus $R_\mathrm{GC}$. In each test, the results remained similar, showing the outcome is not dependent on one unique choice of initial conditions.

We experimented by initially exaggerating the feature to see if the stars would
settle upon the actual model. We began the simulation with a larger dip in the
$\sigma_t$ profile, starting at a value of 50 km s$^{-1}$ within the first 10
kpc instead of 90 km s$^{-1}$. All other conditions remained the same as our
previous simulations. We found that the $\sigma$ profiles flatted and the
anomolous feature dissolved away. We ran the simulations with the same conditions as
previously described, but with a model potential with a perfectly flat rotation
curve, very similar to the one determined in \citet{Kafle2012}. The results
were the same as the other tests. 

Thus we concluded the Milky Way's $\sigma$ and $\beta$ profiles are in a
transient phase, and the observed features at 17~kpc will smooth away within a
few ten million years. 

\section{Summary and Conclusions}
\label{conclusions}
We tested the kinematic model of the Galactic halo proposed by
\citet{Kafle2012}, by directly integrating particles in the \citet{Flynn1996} 3-D model of the Galactic potential, under the assumption of an
initial Gaussian velocity distribution. We found that the particles relaxed quite quickly 
from the initial profiles observed by \citet{Kafle2012}. The features at
17~kpc, most prominently seen in the anisotropy $\beta$ profile as the steep 
decline reaching minimum at $R_\mathrm{GC}=17$~kpc and the sharp rise within a
few kpc, dissolved after 0.02 Gyr. We propose that this feature in the Milky
Way's velocity dispersion $\sigma$ and anisotropy $\beta$ profiles is in a
transient phase, and will flatten within a few ten million years. 

Observational
studies have shown that stellar inhomogeneities and substructures exist in the halo, such
as the Virgo overdensity \citep{Juric2008} and Sagittarius stellar stream
\citep{Ibata1994}, which are presumably remnants of merging
satellites. \citet{Kafle2012} found that masking these two substructures did not
significantly change their modeled profiles, although many underlying stellar
structures yet undetected are likely to exist, such that may or may not disrupt
flat profiles and lead to temporary anomalous features. Such stellar structures
are testable with simulations of halo formation around disc galaxies. \citet{Deason2013} used results from such simulations to show that shell-like structures in the radial velocities of halo stars are tightly related to the furthest orbital distance of stars left over from accreted satellites.

Another
possibility could be that the features in the profiles reflect underlying
properties of the halo. \citet{Carollo2007} found a prograde inner halo and a
retrograde more metal-poor outer halo; and \citet{Carollo2010} and
\citet{de_Jong2010} found the transition between such halo components to lie
between $15<R_\mathrm{GC}/\mathrm{kpc}<20$, which is the location of the
feature in the \citet{Kafle2012} $\sigma$ and $\beta$ profiles. In contrast to
stellar substructures, the presence of a dual halo has been debated for and
against \citep{Schonrich2011,Beers2012,Fermani2013,Schonrich2014}. Despite the
suggested explanations, the origins of (or combination there of) such transient
features in the halo remain unclear.

Alternatively, it is possible that the features in the $\sigma$
  and $\beta$ profiles are due to the broken power-law stellar density
  distribution $\rho$ adopted by \citet{Kafle2012}, as this function
  is marginalised over when determining the stellar velocity
  dispersions. We plan to repeat similar simulations to the current
  paper using the more plausible density of the alpha-beta-gamma
  profile of \citet{Zhao1996}, which is a double power law but
  smooth.

In order to find what causes the features in the Galactic kinematic profiles as
interpreted by \citet{Kafle2012}, we plan to set up a complete library of
dynamically stable halos in the adopted Milky Way potential. The library will be particularlly useful to compare with the large amounts of new data anticipated from surveys such as Gaia and LAMOST.



\section*{Acknowledgements}

S. B. is grateful for the funding provided by the Finnish Cultural Foundation
and the Turku University Foundation and for the hospitality of the Centre for
Astrophysics and Supercomputing at Swinburne University of Technology where the
majority of this research was carried out. S. B. thanks the Emil Aaltonen
Foundation and StarryStory Foundation for travel funding. The authors give special thanks to Prajwal Kafle and HongSheng Zhao for many helpful discussions, comments, and suggestions.

\bibliographystyle{mn2e} 
\bibliography{bibfile} 

\begin{thebibliography}{28}
\expandafter\ifx\csname natexlab\endcsname\relax\def\natexlab#1{#1}\fi

\bibitem[{{Beers} {et~al}\mbox{.}(2012){Beers}, {Carollo}, {Ivezi{\'c}}, {An},
  {Chiba}, {Norris}, {Freeman}, {Lee}, {Munn}, {Re Fiorentin}, {Sivarani},
  {Wilhelm}, {Yanny}, \& {York}}]{Beers2012}
{Beers} T.~C. {et~al.}, 2012, \apj, 746, 34

\bibitem[{{Binney} \& {Tremaine}(2008)}]{Binney2008}
{Binney} J., {Tremaine} S., 2008, {Galactic Dynamics: Second Edition}.
  Princeton University Press

\bibitem[{{Bovy} \& {Rix}(2013)}]{Bovy2013}
{Bovy} J., {Rix} H.-W., 2013, \apj, 779, 115

\bibitem[{{Carollo} {et~al}\mbox{.}(2010){Carollo}, {Beers}, {Chiba}, {Norris},
  {Freeman}, {Lee}, {Ivezi{\'c}}, {Rockosi}, \& {Yanny}}]{Carollo2010}
{Carollo} D. {et~al.}, 2010, \apj, 712, 692

\bibitem[{{Carollo} {et~al}\mbox{.}(2007){Carollo}, {Beers}, {Lee}, {Chiba},
  {Norris}, {Wilhelm}, {Sivarani}, {Marsteller}, {Munn}, {Bailer-Jones},
  {Fiorentin}, \& {York}}]{Carollo2007}
{Carollo} D. {et~al.}, 2007, \nat, 450, 1020

\bibitem[{{de Jong} {et~al}\mbox{.}(2010){de Jong}, {Yanny}, {Rix}, {Dolphin},
  {Martin}, \& {Beers}}]{de_Jong2010}
{de Jong} J.~T.~A., {Yanny} B., {Rix} H.-W., {Dolphin} A.~E., {Martin} N.~F.,
  {Beers} T.~C., 2010, \apj, 714, 663

\bibitem[{{Deason}, {Belokurov} \& {Evans}(2011){Deason}, {Belokurov}, \&
  {Evans}}]{Deason2011}
{Deason} A.~J., {Belokurov} V., {Evans} N.~W., 2011, \mnras, 416, 2903

\bibitem[{{Deason} {et~al}\mbox{.}(2012){Deason}, {Belokurov}, {Evans}, \&
  {An}}]{Deason2012}
{Deason} A.~J., {Belokurov} V., {Evans} N.~W., {An} J., 2012, \mnras, 424, L44

\bibitem[{{Deason} {et~al}\mbox{.}(2013){Deason}, {Belokurov}, {Evans}, \&
  {Johnston}}]{Deason2013}
{Deason} A.~J., {Belokurov} V., {Evans} N.~W., {Johnston} K.~V., 2013, \apj,
  763, 113

\bibitem[{{Deason} {et~al}\mbox{.}(2014){Deason}, {Belokurov}, {Koposov}, \&
  {Rockosi}}]{Deason2014}
{Deason} A.~J., {Belokurov} V., {Koposov} S.~E., {Rockosi} C.~M., 2014, \apj,
  787, 30

\bibitem[{{Fermani} \& {Sch{\"o}nrich}(2013)}]{Fermani2013}
{Fermani} F., {Sch{\"o}nrich} R., 2013, \mnras, 432, 2402

\bibitem[{{Flynn}, {Sommer-Larsen} \& {Christensen}(1996){Flynn},
  {Sommer-Larsen}, \& {Christensen}}]{Flynn1996}
{Flynn} C., {Sommer-Larsen} J., {Christensen} P.~R., 1996, MNRAS, 281, 1027

\bibitem[{{Helmi}(2008)}]{Helmi2008}
{Helmi} A., 2008, \aapr, 15, 145

\bibitem[{{Ibata}, {Gilmore} \& {Irwin}(1994){Ibata}, {Gilmore}, \&
  {Irwin}}]{Ibata1994}
{Ibata} R.~A., {Gilmore} G., {Irwin} M.~J., 1994, \nat, 370, 194

\bibitem[{{Juri{\'c}} {et~al}\mbox{.}(2008){Juri{\'c}}, {Ivezi{\'c}}, {Brooks},
  {Lupton}, {Schlegel}, {Finkbeiner}, {Padmanabhan}, {Bond}, {Sesar},
  {Rockosi}, {Knapp}, {Gunn}, {Sumi}, {Schneider}, {Barentine}, {Brewington},
  {Brinkmann}, {Fukugita}, {Harvanek}, {Kleinman}, {Krzesinski}, {Long},
  {Neilsen}, {Nitta}, {Snedden}, \& {York}}]{Juric2008}
{Juri{\'c}} M. {et~al.}, 2008, \apj, 673, 864

\bibitem[{{Kafle} {et~al}\mbox{.}(2012){Kafle}, {Sharma}, {Lewis}, \&
  {Bland-Hawthorn}}]{Kafle2012}
{Kafle} P.~R., {Sharma} S., {Lewis} G.~F., {Bland-Hawthorn} J., 2012, ApJ, 761,
  98

\bibitem[{{King} {et~al}\mbox{.}(2015){King}, {Brown}, {Geller}, \&
  {Kenyon}}]{KingarXiv2015}
{King}, III C., {Brown} W.~R., {Geller} M.~J., {Kenyon} S.~J., 2015, ArXiv
  e-prints

\bibitem[{{Miyamoto} \& {Nagai}(1975)}]{MiyamotoNagai1975}
{Miyamoto} M., {Nagai} R., 1975, PASJ, 27, 533

\bibitem[{{Press}, {Flannery} \& {Teukolsky}(1986){Press}, {Flannery}, \&
  {Teukolsky}}]{Press1986}
{Press} W.~H., {Flannery} B.~P., {Teukolsky} S.~A., 1986, {Numerical recipes.
  The art of scientific computing}. Cambridge: University Press, pp. 554--560

\bibitem[{{Sandage}(1986)}]{Sandage1986}
{Sandage} A., 1986, \araa, 24, 421

\bibitem[{{Sch{\"o}nrich}, {Asplund} \& {Casagrande}(2011){Sch{\"o}nrich},
  {Asplund}, \& {Casagrande}}]{Schonrich2011}
{Sch{\"o}nrich} R., {Asplund} M., {Casagrande} L., 2011, \mnras, 415, 3807

\bibitem[{{Sch{\"o}nrich}, {Asplund} \& {Casagrande}(2014){Sch{\"o}nrich},
  {Asplund}, \& {Casagrande}}]{Schonrich2014}
{Sch{\"o}nrich} R., {Asplund} M., {Casagrande} L., 2014, \apj, 786, 7

\bibitem[{{Sirko} {et~al}\mbox{.}(2004){Sirko}, {Goodman}, {Knapp},
  {Brinkmann}, {Ivezi{\'c}}, {Knerr}, {Schlegel}, {Schneider}, \&
  {York}}]{Sirko2004}
{Sirko} E. {et~al.}, 2004, \aj, 127, 914

\bibitem[{{Sommer-Larsen} {et~al}\mbox{.}(1997){Sommer-Larsen}, {Beers},
  {Flynn}, {Wilhelm}, \& {Christensen}}]{Sommer-Larsen1997}
{Sommer-Larsen} J., {Beers} T.~C., {Flynn} C., {Wilhelm} R., {Christensen}
  P.~R., 1997, \apj, 481, 775

\bibitem[{{Sommer-Larsen}, {Flynn} \& {Christensen}(1994){Sommer-Larsen},
  {Flynn}, \& {Christensen}}]{Sommer-Larsen1994}
{Sommer-Larsen} J., {Flynn} C., {Christensen} P.~R., 1994, MNRAS, 271, 94

\bibitem[{{Thom} {et~al}\mbox{.}(2005){Thom}, {Flynn}, {Bessell},
  {H{\"a}nninen}, {Beers}, {Christlieb}, {James}, {Holmberg}, \&
  {Gibson}}]{Thom2005}
{Thom} C. {et~al.}, 2005, \mnras, 360, 354

\bibitem[{{Watkins} {et~al}\mbox{.}(2009){Watkins}, {Evans}, {Belokurov},
  {Smith}, {Hewett}, {Bramich}, {Gilmore}, {Irwin}, {Vidrih}, {Wyrzykowski}, \&
  {Zucker}}]{Watkins2009}
{Watkins} L.~L. {et~al.}, 2009, \mnras, 398, 1757

\bibitem[{{Zhao}(1996)}]{Zhao1996}
{Zhao} H., 1996, \mnras, 278, 488

\end{thebibliography}

\label{lastpage}

\end{document}